\documentclass{sig-alternate}
\usepackage{xspace}

\newcommand{\set}[1]{\{#1\}}
\newcommand{\definedas}{\stackrel{def}{=}}
\newcommand{\kron}[1]{\delta\hspace{-0.2111em}\left[#1\right]}

\newcommand{\relret}[1]{}

\newcommand{\query}{q}
\newcommand{\doc}{d}

\newcommand{\corpus}{{\bf {\cal C}}}
\newcommand{\word}{w}
\newcommand{\wordseqlength}{n}
\newcommand{\wordseq}{\word_1 \word_2 \cdots \word_\wordseqlength}
\newcommand{\wordseqvar}{\vec{\word}}
\newcommand{\freq}[2]{f(#1 \in #2)}

\newcommand{\clust}{c}

\newcommand{\cluster}{\clust}

\newcommand{\nbhd}[1]{\mathrm{Cohort}(#1)} 
\newcommand{\nbhdEng}{cohort}
\newcommand{\clusters}[1]{{\rm Clusters}(#1)} 

\newcommand{\sizeclust}{k} 
\newcommand{\facetsfull}[3]{{\rm Facets}_{#2}(#1)} 
\newcommand{\facets}{\facetsfull{\doc}{\query}{\corpus}}

\newcommand{\prob}{p}

\newcommand{\ilmprob}{\prob} 

\newcommand{\condArbP}[3]{\ensuremath{#1(#2 \vert #3)}}
\newcommand{\condP}[2]{\condArbP{\prob}{#1}{#2}}

\newcommand{\inducedprob}[3]{\ensuremath{#1_{#2}(#3)}}
\newcommand{\inducedlmprob}[2]{\inducedprob{\ilmprob}{#1}{#2}}
\newcommand{\mlprob}[2]{\inducedprob{\ilmprob^{ML}}{#1}{#2}}
\newcommand{\docinducedlmprob}[1]{\inducedlmprob{\doc}{#1}}
\newcommand{\docinducedlmprobmulti}[1]{\inducedprob{\ilmprob^{Dir}}{\doc}{#1}}
\newcommand{\clustinducedlmprob}[1]{\inducedlmprob{\cluster}{#1}}

\newcommand{\numretclust}{m} 

\newcommand{\priorconst}{\alpha}
\newcommand{\numretdocs}{N}
\newcommand{\topdocsfull}[3]{{\rm TopDocs}(#3)}
\newcommand{\topdocs}{\topdocsfull{\clusters{\corpus}}{\query}{\numretdocs}}
\newcommand{\topclustsfull}[3]{{\rm TopClusters}_{#2}(#3)}
\newcommand{\topclusts}{\topclustsfull{\clusters{\corpus}}{\query}{\numretclust}}
\newcommand{\topclustsrerank}{\topclusts}

\newcommand{\topclustsaspect}{\topclusts}

\newcommand{\firstmention}[1]{{\bf #1}}

\newcommand{\scorererank}{select}

\newcommand{\baseline}{LM}

\newcommand{\centroidAlg}{basis-\scorererank\xspace}
\newcommand{\abbrevcentroidAlg}{basis-S\xspace}

\newcommand{\fifoterm}{set}

\newcommand{\fifoAlg}{\fifoterm-\scorererank\xspace}
\newcommand{\abbrevfifoAlg}{set-S\xspace}

\newcommand{\doubleCounted}{bag-\scorererank\xspace}
\newcommand{\abbrevdoubleCounted}{bag-S\xspace}

\newcommand{\aspectterm}{aspect-x\xspace}
\newcommand{\Aspectterm}{Aspect-x\xspace}

\newcommand{\mrr}{{\aspectterm}\xspace}
\newcommand{\abbrevmrr}{{\aspectterm}\xspace}
\newcommand{\MicroAlg}{Interpolation\xspace}
\newcommand{\microAlg}{interpolation\xspace}
\newcommand{\abbrevmicroAlg}{interp.\xspace}
\newcommand{\doubleCountAspect}{uniform-{\aspectterm}\xspace}
\newcommand{\abbrevdoubleCountAspect}{uniform\xspace}

\newcommand{\groupSize}[1]{\left\vert #1 \right\vert}

\numberofauthors{2}
\author{
\alignauthor Oren Kurland and Lillian Lee\\
\affaddr{Department of Computer Science}\\
\affaddr{Cornell University}\\
\affaddr{Ithaca, NY 14853-7501}\\
\email{\footnotesize \{kurland,llee\}@cs.cornell.edu}
}
\title{Corpus Structure, Language Models, and Ad Hoc Information Retrieval}

\begin{document}
\conferenceinfo{SIGIR'04,}{July 25--29,2004, Sheffield, South Yorkshire, UK.}
\CopyrightYear{2004}
\crdata{1-58113-881-4/04/0007}

\maketitle

\begin{abstract}

Most 
previous work on the recently developed {\em language-modeling}
approach to information retrieval focuses on 
docu\-ment-specific
characteristics, 
and therefore does not take into account the structure of the surrounding corpus.
We propose a novel algorithmic framework in which  information
provided by document-based language models is enhanced by the incorporation of information drawn from 
{\em clusters} of similar documents.
Using this framework, we develop a suite of new algorithms. Even the
simplest 
typically outperforms the standard language-modeling approach in precision and
recall, 
and our new {\em interpolation} algorithm  posts statistically significant improvements for
both metrics over all three corpora tested.
\end{abstract}

\category{H3.3}{Information Search and Retrieval}{Language models, clustering, smoothing}
\terms{Algorithms, Experiments}

\section{Introduction}
\label{sec:intro}

As is well known, a basic problem in information retrieval is to
determine how relevant a particular document is to a query.  In the
automatic ad hoc retrieval setting, examples of relevant documents are
not supplied.  Given this absence of explicit relevance evidence,
it is important to consider what other information sources 
can be exploited.

In methods patterned after the classic tf.idf 
document-vector
approach to text
representation, the focus is mostly on utilizing within-document
features, such as term frequencies.  Information drawn from the corpus
as a whole generally consists of aggregates of statistics gathered
from each document considered in isolation; for example, the inverse
document frequency is based on checking, for each document, whether
that document contains a particular term.

Recent work has demonstrated the effectiveness of an alternative 
approach
wherein
probabilistic models of text generation are constructed from documents, and 
these
induced {\em language models} (LMs) are used to perform document
ranking \cite{Ponte+Croft:98a,Croft+Lafferty:03a}. 
Like tf.idf and related techniques, though, 
lang\-uage-modeling methods typically use only in\-di\-vi\-dual-doc\-u\-ment 
features and corpus-wide aggregates of the same.
(Corpus term counts are generally employed
for {\em smoothing}, so that unseen text can be  assigned non-zero
probability.)

Neither of the aforementioned approaches typically makes use of a
potentially very powerful source of information: the {\em similarity
structure} of the corpus.   
Clusters are a convenient representation of similarity whose potential
for improving retrieval performance has long been recognized
\cite{Croft:80a,vanR:79a}.
From our point of view, one key
advantage is that they provide smoothed, more representative statistics for
their elements, as has been recognized in statistical natural language
processing for some time \cite{Brown+al:90a}. For example, we could
infer that a document not containing a certain query term is still
relevant if the document belongs to a cluster whose component
documents generally do contain the term.

However, relying on clusters alone has some potential drawbacks.  
Clustering at retrieval time can be very expensive, but off-line
clustering seems, by definition, query-independent and therefore may be
based on factors that are irrelevant to user information need.  Also,
cluster statistics may over-generalize with respect to specific member
documents.

We therefore propose a framework for incorporating both corpus-structure
information ---  using pre-computed, {overlapping} clusters ---  and indi\-vidual-document
information.
Importantly, although  cluster formation is query-independent, within
our framework the
{\em choice} of which clusters to incorporate 
{\em can} depend on
the query.  We then consider several of the many possible
algorithms arising as specific instantiations of 
our framework.  These
include
both novel methods and, as special cases, both the standard,
non-cluster-based LM approach and 
a variant of the 
cluster-based aspect model \cite{Hofmann+Puzicha:98b}.

Our empirical evaluation consists of experiments in an array of
settings created by varying several parameters and meta-param\-eters;
these include the corpus, the information representation (e.g.,
language models versus tf.idf-style vectors), and, when applicable,
the smoothing method selected.  We find that even the worst-performing
of our novel algorithms is competitive with the LM approach, and
indeed always provides substantial improvement in recall.  In general,
our algorithms provide good performance in comparison to a number of
recently proposed methods,  thus demonstrating
that our integration approach to incorporating document and
corpus-structure information is an effective way to improve ad hoc
retrieval.

\paragraph*{Notational conventions}

We use $\doc,\query, \cluster$ and $\corpus$ to denote a 
document, query, cluster, and corpus, respectively.  A fixed
vocabulary is assumed.  We use the notation
$\docinducedlmprob{\cdot}$ for the {\em language model} --- which
assigns probabilities to text strings over the fixed vocabulary ---
induced  from  $\doc$ by some pre-specified method, and
$\clustinducedlmprob{\cdot}$ for the language model induced from
$\cluster$. (Section \ref{sec:lmbasics} describes the induction
methods we used in our experiments.)

It is convenient to use {\em Kronecker delta} notation $\kron{s}$
to set up some definitions.  The argument $s$
is a statement; $\kron{s} = 1$ if $s$ holds, 0 otherwise.

\section{Retrieval Framework}
\label{sec:structure}

As noted above, when we rank documents with respect to a query, we
desire per-document scores that rely both on information drawn
from the particular document's contents and on how the document is situated
within the similarity structure of the ambient corpus.  

\paragraph*{Structure representation via overlapping clusters}
Document clusters are an attractive choice for representing corpus
similarity structure (see \cite[chapter 3]{vanR:79a} for extended
discussion).
Clusters  can be thought of as 
{\em facets} of the corpus that users might be interested in.
Given that a
particular document can be relevant to a 
user for several
reasons, or to different users for different reasons,
we believe 
that a set of {overlapping}
clusters\footnote{We include soft or probabilistic clusters in this
category.}  
forms a better model for similarity structure than 
a partitioning of the corpus.  Furthermore, employing intersecting clusters 
may reduce
information loss due to the generalization that clustering can
introduce
\cite[pg. 44]{vanR:79a}.

\paragraph*{Information representation}

Motivated by the empirical successes of language-modeling-based
approaches \cite{Ponte+Croft:98a,Croft+Lafferty:03a}, we use language
models induced from documents and clusters 
as our information representation.  Thus,
$\docinducedlmprob{\query}$ and $\clustinducedlmprob{\query}$ specify
our initial knowledge of the relation between the query $\query$ and a
particular document $\doc$ or cluster $\cluster$, respectively.
(However, Section \ref{sec:results} shows that using a tf.idf
representation  also yields performance improvements with respect to
the appropriate baseline, though not to the same degree as using
language models does.)

\paragraph*{Information integration}

To assign a ranking to the documents in a corpus $\corpus$ with
respect to  $\query$, we want to score each $\doc \in
\corpus$  against $\query$ in a way that  incorporates information
from query-relevant corpus facets to which $\doc$ belongs. While one could compute clusters specific to $\query$ at retrieval
time, efficiency considerations 
compel us to create $\clusters{\corpus}$, the set of
clusters,  in advance, and hence
in a query-independent fashion.  
To compensate, at retrieval time we 
base the {\em choice} of appropriate facets on the query.  

How might cluster information be used?  Our discussion above indicates
that clusters can serve two roles.  Insofar as they approximate true
facets 
of the corpus, they can aid in the {\em selection} of relevant
documents: we would want to retrieve those that belong to clusters
corresponding to facets of interest to the user.  On the other hand,
clusters also have the capacity to {\em
smooth} individual-document language models, since they pool 
statistics from multiple documents.  Finally, we must
remember that over-reliance on $\clustinducedlmprob{\query}$ can
over-generalize by failing to account for  document-specific
information encoded in $\docinducedlmprob{\query}$.

These observations motivate the algorithm template shown in Figure
\ref{fig:alg-template}.  This template is fairly general:  both the standard language-modeling approach \cite{Ponte+Croft:98a} and the
aspect model
\cite{Hofmann+Puzicha:98b} are concrete instantiations.  In the
template, the choice of $\facets$ corresponds to utilizing clusters in
their selection
role.  The scoring step can be thought of as integrating 
$\docinducedlmprob{\query}$ with cluster-based language models in
their smoothing role.  The optional re-ranking step is used as a way
to further bias the final ranking towards document-specific information, if
desired. 
 Note that re-ranking can change the average non-interpolated
 precision but not the absolute precision or recall of the retrieval
 results;
we therefore use it, when necessary, to enhance average precision.
(Section \ref{sec:results} reports experiments studying its
efficacy.)

\begin{figure}[ht]
\fbox{
\begin{minipage}{3.3in}
\begin{tabbing}
Sco\=rin\=g t\=ime: G\=iven $\query$ and $\doc \in \corpus$\kill
{\em Offline}:  Create $\clusters{\corpus}$
\\
{\em Given $\query$ and $\numretdocs$, the number of documents to retrieve}:\\
\> For each  $\doc \in \corpus$,\\
\>	  \> \>Choose a  cluster subset $\facets \subseteq \clusters{\corpus}$\\
\>	  \> \>Score $\doc$ by a weighted combination of $\docinducedlmprob{\query}$ and  \\ 
\>	  \> \>	\> the $\clustinducedlmprob{\query}$'s for all
$\cluster \in \facets$ \\
\> Set $\topdocs$ to the  rank-ordered list of $\numretdocs$ top-\\
\>    \>  scoring documents \\
\> Optional: re-rank 
$\doc \in \topdocs$ by $\docinducedlmprob{\query}$\\
\> Return $\topdocs$ \\
\end{tabbing}
\end{minipage}
} 
\caption{\label{fig:alg-template} Algorithm template. 
}
\end{figure}

In the next section, we describe a number of specific algorithms
arising from this template, concentrating on their degree
of dependence  on  cluster-induced language models.

\section{Retrieval Algorithms}
\label{sec:retrievalAlgo}

\begin{table*}[t]
\begin{tabular}{|l*{3}{c}|}\hline
  & $\facets$ & Score & Re-rank by \docinducedlmprob{\query}? \\ 
\hline\hline
\multicolumn{1}{|l|}{ \baseline} & { N/A} & { $\docinducedlmprob{\query}$} & { (redundant)} \\ 
\hline
\multicolumn{1}{|l|}{ \centroidAlg} & { $\set{\nbhd{\doc}} \cap \topclustsrerank$} & { $\docinducedlmprob{\query} \cdot \kron{\groupSize{\facets}>0}$} & { (redundant)} \\ 
\multicolumn{1}{|l|}{ \fifoAlg} & { $\set{\cluster: \doc \in \cluster} \cap \topclustsrerank$} & { $\docinducedlmprob{\query} \cdot \kron{\groupSize{\facets}>0}$} & { (redundant)} \\ 
\multicolumn{1}{|l|}{ \doubleCounted} & { $\set{\cluster: \doc \in \cluster} \cap \topclustsrerank$} & { $\docinducedlmprob{\query} \cdot \groupSize{\facets}$} & { yes} \\ 
\hline
\multicolumn{1}{|l|}{ \doubleCountAspect} & { $\set{\cluster: \doc \in \cluster} \cap \topclustsaspect$} & { $\sum_{\cluster \in \facets} \clustinducedlmprob{\query}$} & { yes} \\ 
\multicolumn{1}{|l|}{ \mrr} & { $\set{\cluster: \doc \in \cluster} \cap \topclustsaspect$} & { $\sum_{\cluster \in \facets} \clustinducedlmprob{\query}\cdot \clustinducedlmprob{\doc}$} & { yes} \\ 
\hline
\multicolumn{1}{|l|}{ \microAlg} & { $\set{\cluster: \doc \in \cluster} \cap \topclustsaspect$} & { $\lambda \cdot \docinducedlmprob{\query} +  (1 - \lambda) \sum_{\cluster \in \facets}  \clustinducedlmprob{\query}  \cdot \clustinducedlmprob{\doc}$} & { no} \\ 
\hline
\end{tabular}

\caption{\label{tab:models} Algorithm specifications.}
\end{table*}

Table \ref{tab:models} 
summarizes the algorithms we consider, which
represent a few choices out of
the many possible ways to instantiate the template of
Figure \ref{fig:alg-template}. 
Our preference in picking these
algorithms has been towards simpler methods, so as to
focus on the impact
of using cluster information (as opposed to the impact of tuning many
weighting parameters).

\paragraph*{First step: Cluster formation and selection}

There are many  algorithms that can be used to create $\clusters{\corpus}$,
the set of overlapping document clusters 
 required by 
 Figure \ref{fig:alg-template}'s 
template. 
In 
our experiments, we simply have each
document $\doc$ form the {\em basis} of a cluster $\nbhd{\doc}$ consisting of $\doc$
and its $\sizeclust - 1$ nearest neighbors,
where $\sizeclust$ is a free parameter.  (Note that two clusters with
different basis documents may contain the same set of documents.)
Inter-document distance is measured by the Kullback-Leibler (KL) divergence
between the corresponding (smoothed) language models,
as in
\cite{Lafferty+Zhai:01a}.

The
idea  behind our use of {\nbhdEng}s is that a document's nearest neighbors in similarity space
represent a local ``fragment'' or ``tile'' of the overall similarity
structure of the corpus.  
Our evaluation results show that even this relatively
unsophisticated way to approximate facets enables effective
leveraging of corpus structure; at the very least, it serves as a form
of nearest-neighbor smoothing (see below).

The first retrieval-time action specified by our algorithm template is
to choose $\facets$, a query-dependent subset of $\clusters{\corpus}$.
In all the algorithms described below except the baseline (which
doesn't use cluster information), there is a document-selection
aspect to this subset, in that only documents in some $\cluster \in
\facets$ can appear in the final ranked-list output.
Ideally, we would use the clusters best approximating those (true) facets of
the corpus that are most representative of the user's interests, as expressed by
$\query$;  therefore,  we require that $\facets$ be a subset of $\topclusts$, the
top $\numretclust$ clusters $\cluster$ with respect to
$\clustinducedlmprob{\query}$.
But we also want to evaluate $\doc$ only with respect to the facets it actually exhibits.  Thus,
in what follows (except for the baseline),
$\facets$ is always defined to be a subset of $\set{\cluster : \doc
  \in \cluster} \cap \topclusts$; we assume $\numretclust$ is large
enough to produce the desired number of retrieved documents $\numretdocs$.

\paragraph*{Baseline method}  The {baseline}  for 
our
experiments, denoted  \firstmention{\baseline}, is 
to simply rank documents 
by
$\docinducedlmprob{\query}$ --- no cluster information is
used. Details of our particular implementation are  given in Section
\ref{sec:experiments}.

\paragraph*{Selection methods}  In this class of algorithms, the cluster-induced language models play a very
small role once the set $\facets$ is selected.
In essence, the standard language-modeling approach
(that is, ranking by $\docinducedlmprob{\query}$) is invoked to rank (some
of) the documents comprising the clusters in $\facets$.  This 
method of scoring 
is
intended to serve as a precision-enhancing mechanism, 
downgrading documents that happen to be members of some 
$\cluster \in \facets$ by dint of similarity 
to $\doc$ 
in respects not pertinent
to $\query$.

In the \firstmention{\centroidAlg} algorithm, the net effect of the
definition given in Table \ref{tab:models} is that only the {\em basis}
documents of the clusters in $\topclusts$ are allowed to 
appear
in the final output list.  Thus, this algorithm uses the
pooling of statistics from documents in $\nbhd{\doc}$ simply to decide
whether $\doc$ is worth ranking;
the rank itself is based solely on $\docinducedlmprob{\query}$.

The \firstmention{\fifoAlg} algorithm differs  in
that {\em all} the documents in the clusters in $\topclusts$ may
appear in the final output list --- the ``set'' referred to in the
name is the union of the clusters in $\topclusts$.  The idea is that any
document in a ``best'' cluster, basis or not, is potentially relevant
and should be ranked.  Again, the ranking of the selected documents is
by $\docinducedlmprob{\query}$.\footnote{
Because our implementation treats clusters and their
component documents in a ``fifo''
manner, 
it deviates
slightly from the template.
Let $\numretdocs'$ be the number of documents in the 
  $\numretclust - 1$ highest-ranked clusters.
Then, only the $\numretdocs - \numretdocs'$ documents in the  $\numretclust$'th
 cluster that are closest, in the KL-divergence sense, 
to the
cluster's basis
are allowed into $\topdocs$.

}

Another natural variant of the same idea is that documents appearing
in more than one cluster in $\topclusts$ should get extra
consideration, given that they 
appear in several (approximations of) facets
thought to be of interest to the user.  This idea gives rise to the
\firstmention{\doubleCounted} algorithm,
so named in reference to the incorporation of a document's
multiplicity in the {\em bag} formed from the ``multi-set union'' of
all the clusters in $\facets$.
First,  each
selected document $\doc$ is assigned a score consisting of the product
of its language-modeling score
$\docinducedlmprob{\query}$ and the number of ``top''
clusters it belongs to.  The  $\numretdocs$ top-scoring documents are then
re-ranked via $\docinducedlmprob{\query}$ and presented in the new
sorted order.

\paragraph*{\Aspectterm methods}  
We now turn to algorithms making more
explicit use of clusters as smoothing mechanisms.  In particular, we
study what we  term ``\aspectterm'' methods.  Our choice of name is a  reference to the work of
Hofmann and Puzicha \shortcite{Hofmann+Puzicha:98b},  which conceives of
clusters as explanatory latent variables underlying the observed
data. 
(The ``x'' stands for ``extended'').
In our setting, this idea translates to using
$\clustinducedlmprob{\query}$ as a proxy for
$\docinducedlmprob{\query}$, where the degree of dependence on a
particular $\clustinducedlmprob{\query}$ is based on the strength of 
association between $\doc$ and $\cluster$.  The \firstmention{\mrr}
algorithm measures this association by $\clustinducedlmprob{\doc}$;
the \firstmention{\doubleCountAspect} algorithm assumes that every $\doc \in
\cluster$ has the same degree of association to $\cluster$.  
In both cases, re-ranking by $\docinducedlmprob{\query}$ is
applied.

The scoring function we use for 
our \mrr algorithm can be motivated by
appealing to the 
probabilistic derivation of the aspect
model \cite{Hofmann+Puzicha:98b}, as follows. It is a fact that 
\begin{eqnarray}
\condP{\query}{\doc} & = &  \sum_\cluster
\condP{\query}{\doc,\cluster}\condP{\cluster}{\doc}. \label{eqn:factor}
\end{eqnarray}
The {aspect model}  assumes that a query
is conditionally independent of a document given a cluster (which is a
way of using clusters to smooth individual-document statistics), in which
case $\condP{\query}{\doc} = 
\sum_\cluster\condP{\query}{\cluster}\condP{\cluster}{\doc}$.  If we
further assume that $\prob(\doc)$ and $\prob(\cluster)$ are
constant, we can  write
$\condP{\query}{\doc} =
\priorconst\sum_\cluster \condP{\query}{\cluster}\condP{\doc}{\cluster}$, 
where $\priorconst$ is a constant that doesn't affect ranking.
Our \mrr algorithm then arises by replacing the conditional probabilities with the corresponding
language models and only summing over the clusters in $\facets$.
Constraining which clusters participate in the sum to those of
relatively high rank is
important:  
experiments indicate that using a large number of
clusters could be detrimental.
We note, however,  that it appears difficult within the strictly probabilistic
framework of the original aspect model  to
incorporate such a constraint: a particular cluster's rank depends on
all the other clusters, but none of the terms in the basic
aspect-model equation 
explicitly conditions on them.

\paragraph*{A hybrid algorithm}  The selection-only algorithms emphasize $\docinducedlmprob{\query}$ in
scoring a document $\doc$; in contrast,  the
\aspectterm algorithms rely on $\clustinducedlmprob{\query}$.
We created the \firstmention{\microAlg}
algorithm to combine the advantages of these two
approaches.

The algorithm can be derived by dropping the original aspect model's
 conditional independence assumption --- namely, that
 $\condP{\query}{\doc,\cluster} = \condP{\query}{\cluster}$ --- and
instead setting 
$\condP{\query}{\doc,\cluster}$ in Equation
\ref{eqn:factor} to $\lambda  \condP{\query}{\doc} + (1 - \lambda)
 \condP{\query}{\cluster}$, where $\lambda$
indicates 
the degree of emphasis on individual-document 
information.  If we do so, then via some algebra we get
$\condP{\query}{\doc} = \lambda \condP{\query}{\doc} + (1 -
\lambda) \sum_{\cluster}
\condP{\query}{\cluster}\condP{\cluster}{\doc}$. Finally, 
applying the
same assumptions as described in our discussion of the \mrr algorithm
yields a score function that is the linear interpolation of the
score of the standard LM approach and the score of the \mrr algorithm.
Note that no re-ranking step occurs; as 
we shall see, the \microAlg
algorithm's
incorporation of document-specific information 
yields
higher precision.

\section{Related Work}
\label{sec:relwork}

Document clustering has a long history in information retrieval
\cite{Croft:80a,vanR:79a}; in particular, approximating topics via clusters is a
recurring theme \cite{Xu+Croft:99a}.  
Arguably the work most related to ours by dint of
employing both clustering and language modeling in the context of ad
hoc retrieval\footnote{
See e.g., \cite{Brown+al:90a},
\cite{Iyer+Ostendorf:99a}, and \cite{Hearst+Pedersen:96a} for
applications of clustering in 
related areas.} is that on latent-variable models, e.g.,
\cite{Hofmann+Puzicha:98b,Hofmann:01a,Lavrenko:02a,Blei+Ng+Jordan:03a}, of which the
classic aspect model is one instantiation.  Such work takes a strictly
probabilistic approach to the problems we have discussed with standard
language modeling, as opposed to our algorithmic viewpoint.  Also, a
focus in the latent-variable work has been on sophisticated cluster
induction, whereas we find that a very simple clustering scheme works
rather well in practice.  
Interestingly, Hofmann \cite{Hofmann:01a}
linearly interpolated his probabilistic model's score, which is based on (soft)
clusters, with the usual cosine 
metric;
this is quite close in spirit to what
our \microAlg algorithm does.

Implicit corpus structure is also exploited 
by Lafferty and Zhai's {\em expanded query language model}
\cite{Lafferty+Zhai:01a}.
Their method 
uses interleaved 
document-term Markov chains (which can be
thought of as tracing ``paths'' between related documents)
to 
enhance
language models built from queries.  This is similar conceptually
to our framework's use of inter-document similarities 
to enhance the performance of document language models, although 
in our work the
notion of similarity is more explicit.

\medskip

\newcommand{\better}[1]{\mathit{#1}}
\newcommand\relone{}
\begin{table*}[t]
\begin{flushleft}
\small
\centering
\begin{tabular}{|l||c||c|c|c|c|c|c||c|}
\hline
&Baseline: \baseline&\abbrevcentroidAlg & \abbrevfifoAlg &  
\abbrevdoubleCounted & \abbrevdoubleCountAspect &\abbrevmrr &
\abbrevmicroAlg & Pseudo-feedback Markov chains\\  \hline \hline
Avg. Prec.& $21.03\%$ & $\better{22.1\%^*}$ & $\better{22.45\%^*}$ & $\better{22.3\%^*}$ & $\better{21.7\%}$  & $\better{22.6\%^*}$ &\mbox{\boldmath$24.9\%^*$} & $\better{23.2\%}$\\ \hline
Prec. at 0 &$57.4\%$ & \mbox{\boldmath$58.5\%$} & \mbox{\boldmath$58.5\%$} & $\better{58.1\%}$& $57.2\%$ & $\better{58.2\%}$ & $55.8\%$ & $53.4\%$\\ \hline
\relret{RelRet & $1587\relone$ & $\better{1789\relone}$ &  $\better{1831\relone}$ & $\better{2047\relone}$ & $\better{1869\relone}$ & $\better{2027\relone}$& \mbox{\boldmath$2065\relone$} & $\better{2019\relone}$ \\ \hline}
Recall & $48.67\%$ & $\better{54.86\%}$  & $\better{56.15\%}$ & $\better{62.77\%^*}$& $\better{57.31\%}$  & $\better{62.16\%^*}$& \mbox{\boldmath$63.62\%^*$} & $\better{61.91\%}$\\ \hline
\end{tabular}
\caption{\label{tab:AP89} AP89 results (3261 relevant documents). Cluster size $\sizeclust=40$; interpolation parameter $\lambda=0.4$.}
\end{flushleft}
\end{table*}

\newcommand{\reltwo}{}
\begin{table*}[th]
\begin{flushleft}
\small
\centering
\begin{tabular}{|l||c||c|c|c|c|c|c||c|}
\hline
&Baseline: \baseline& \abbrevcentroidAlg &\abbrevfifoAlg & \abbrevdoubleCounted & \abbrevdoubleCountAspect & \abbrevmrr & \abbrevmicroAlg & Relevance model \\  \hline \hline
Avg. Prec.& $24.37\%$ &  $\better{26.58\%^*}$ & $\better{28.11\%^*}$ & $\better{26.65\%^*}$ & $\better{24.92\%}$ &$\better{27.5\%^*}$ & \mbox{\boldmath$31.28\%^*$} & $\better{26.17\%}$ \\ \hline
Prec. at $0$ & $65.52\%$ & $\better{65.77\%}$ & $65.32\%$ & $65.17\%$ & \mbox{\boldmath$67.5\%$} & $\better{65.9\%}$ & $\better{65.82\%}$ & $61.61\%$ \\ \hline
\relret{RelRet & $3197\reltwo$ & $\better{3453\reltwo}$ &  $\better{3675\reltwo}$ & $\better{3807\reltwo}$ & $ \better{3318\reltwo}$& $\better{3791\reltwo}$ & \mbox{\boldmath$3884\reltwo$} & $\better{3733\reltwo}$ \\ \hline}
Recall & $66.53\%$ & $\better{71.86\%}$ & $\better{76.48\%^*}$ & $\better{79.23\%^*}$ & $\better{69.05\%}$ & $\better{78.9\%^*}$ & \mbox{\boldmath$80.83\%^*$} & $\better{77.69\%}$\\ \hline
\end{tabular}
\caption{\label{tab:AP88+89} AP88+89 results (4805 relevant documents).  Cluster size $\sizeclust=40$; interpolation parameter $\lambda=0.6$.}
\end{flushleft}
\end{table*}

\newcommand{\relthree}{}

\begin{table*}[th]
\begin{flushleft}

\small
\centering
\begin{tabular}{|l||c||c|c|c|c|c|c|}
\hline
 &Baseline: \baseline& \abbrevcentroidAlg & \abbrevfifoAlg & \abbrevdoubleCounted & \abbrevdoubleCountAspect & \abbrevmrr & \abbrevmicroAlg \\  \hline \hline
Avg. Prec.& $22.16\%$ & $21.92\%$ & $\better{22.52\%}$ & $22\%$ & $21.73\%$ & $\better{22.45\%}$ & \mbox{\boldmath$23.88\%^*$} \\ \hline
Prec. at $0$ & $57.37\%$ & $\better{57.91\%}$ & $\better{57.89\%}$ & $\better{58.16\%}$ & $57.28\%$  & \mbox{\boldmath$58.25\%$} & $\better{57.81\%}$ \\ \hline
\relret{RelRet & $672\relthree$ & $\better{685\relthree}$ & $\better{774\relthree}$ &  $\better{808\relthree}$ &  $\better{742\relthree}$ & $\better{808\relthree}$& \mbox{\boldmath$886\relthree$} & $\better{851\relthree}$ \\ \hline}
Recall &$48.31\%$&$\better{55.64\%}$ & $\better{58.09\%}$ & $\better{53.34\%}$ & $\better{58.09\%}$& \mbox{\boldmath$63.7\%^*$} & $\better{61.18\%^*}$ \\ \hline

\end{tabular}
\caption{\label{tab:LA+FR} Results for LA+FR (1391 relevant documents).
Cluster size $\sizeclust=10$; interpolation parameter 
 $\lambda=0.8$.}
\end{flushleft}
\end{table*}

\begin{figure*}[t]
\newcommand{\mytabwidth}{2.3in}
\begin{tabular}{ccc}
{\bf AP89} & {\bf AP88+89} & {\bf LA+FR} \\
\epsfig{file= 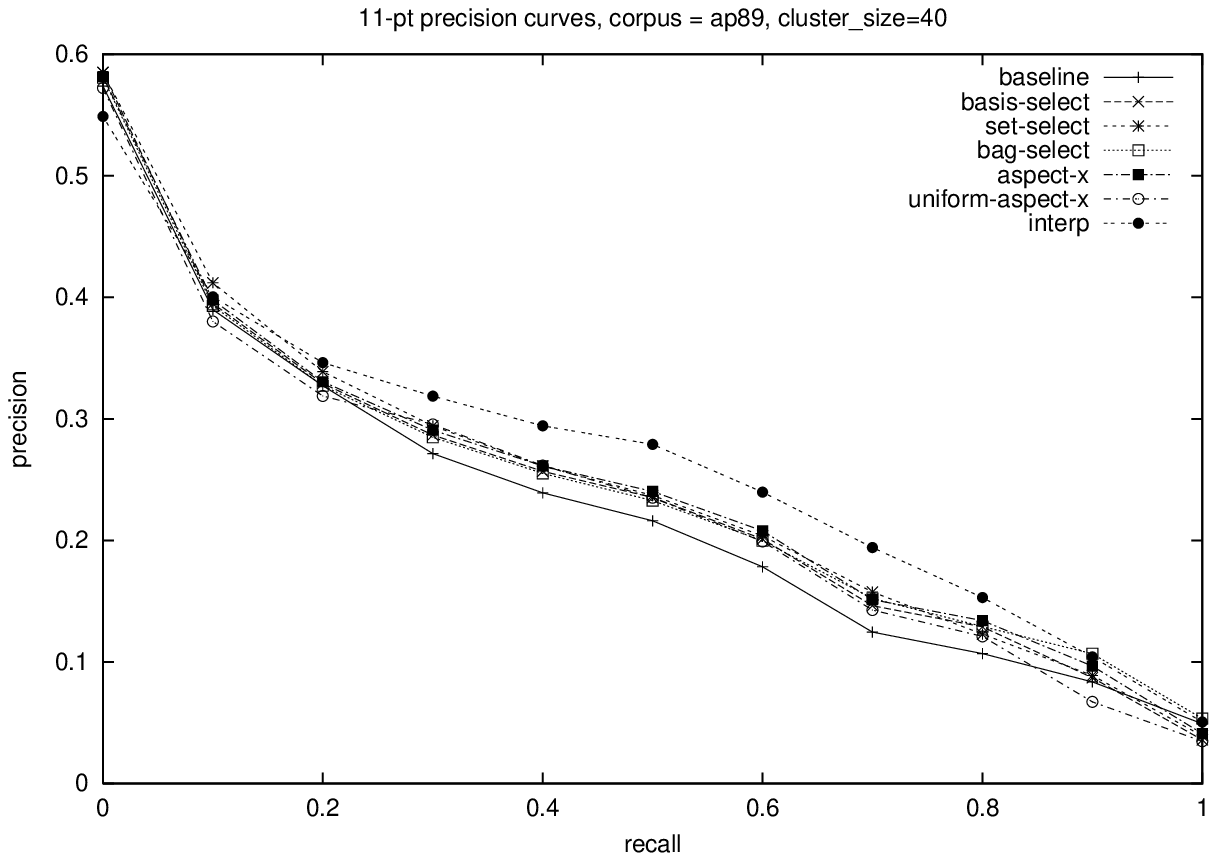,
  width= \mytabwidth}
&
\epsfig{file=
  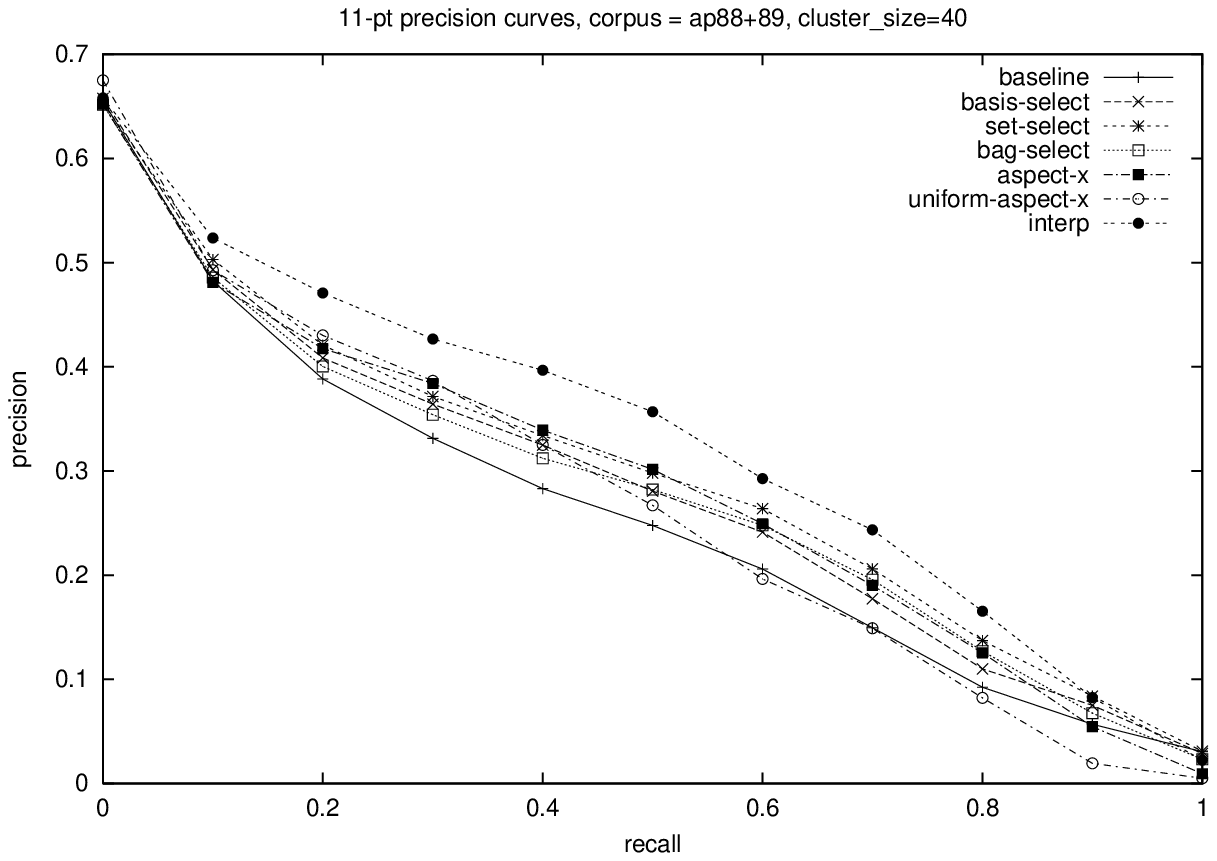, width=
  \mytabwidth} 
&
\epsfig{file=
  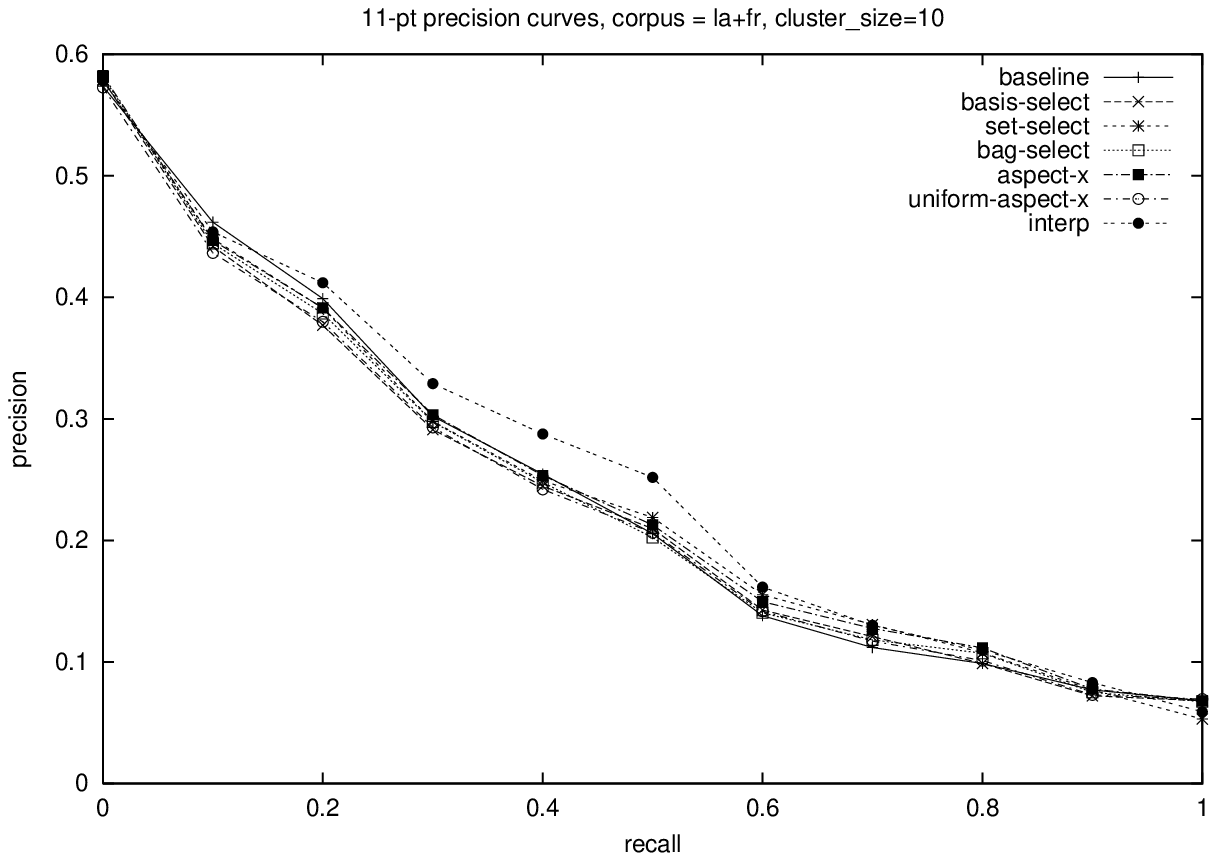, width=
  \mytabwidth }
\end{tabular}
\caption{\label{fig:interPrec} 11-point precision/recall curves.  For
  AP89 and AP88+89, $\sizeclust=40$; for LA+FR, $\sizeclust=10$.}
\end{figure*}

\section{Experimental Setup}
\label{sec:experiments}

\paragraph*{Data}
We conducted our experiments on TREC
data. We used titles (rather than  full
descriptions) as queries, resulting in an average length of 2-5
terms. Some characteristics of our three corpora are summarized
in the following table.
\begin{center}
\begin{tabular}{|lrcl|}\hline
corpus & \# of docs & queries & previous work \\ \hline
AP89 & 84,678 & 1-46,48-50 & Lafferty \& Zhai \cite{Lafferty+Zhai:01a}\\
AP88+89 & 164,597 & 101-150 & Lavrenko  \& Croft  \cite{Lavrenko+Croft:01a}\\
LA+FR  & 187,526 & 401-450 & $\quad \quad \quad$- \\  \hline
\end{tabular}
\end{center}
The first two data corpora, AP89 and AP88+89, were chosen because they
have served as data
for previous research on state-of-the-art 
algorithms somewhat related to but considerably extending the basic LM approach.
We used the 
same
stemming and stopword-removal
policies 
as in those previous experiments;
  hence, 
we applied the Porter stemmer to  the AP89 collection (disk one),
and  
we ran the Krovetz stemmer on AP88+89 and removed 
both INQUERY stopwords
\cite{Allan+EtAl:2001} and length-one tokens.
LA+FR (disk 5 and  4, respectively), which is part of the TREC-8 corpus
(we 
 used TREC-8 ad hoc queries), 
was neither stemmed
nor subjected to stopword removal. 
This corpus is more heterogeneous
than the other two.  

\paragraph*{Induction of base language models}
\label{sec:lmbasics}

Unless otherwise specified, we use unigram Dirichlet-smoothed language
models (which were previously shown to yield good performance for
short queries
\cite{Zhai+Lafferty:01a}) in the following manner.
For the purposes of this discussion, we use the term ``document'' and notation $\doc$ to refer either to a
true document in the corpus $\corpus$ or to a query.
Let $\freq{x}{y}$ be the
number of times word $x$ occurs in item $y$. For
a text sequence $\wordseqvar = \wordseq$,
the Dirichlet-smoothed
language model induced from  $\doc$ assigns the following
probability to $\wordseqvar$:

\begin{equation*}
\docinducedlmprobmulti{\wordseqvar} \definedas \prod_{i=1}^{\wordseqlength}
\frac{\freq{\word_i}{\doc} + \mu \cdot
\mlprob{\corpus}{\word_i}}
{\sum_{\word} \freq{\word}{\doc} + \mu }\,,
\end{equation*}
where the free parameter  $\mu$ controls the degree to which the document's
statistics are altered by the overall corpus statistics, and 
``$ML$'' indicates the maximum-likelihood estimate. Then, for any two
documents $\doc$ and $\doc'$,
we set $\docinducedlmprob{\doc'}$ to 
$$\exp\left(-D\left(\mlprob{\doc'}{\cdot} \vert
\vert\docinducedlmprobmulti{\cdot}\right)\right)$$ (normalizing when
appropriate), where $D$ is the
Kullback-Leibler divergence.  This formulation is actually equivalent
to a log-likelihood criterion  under certain
assumptions \cite{Lafferty+Zhai:01a}, but in practice is less sensitive than
\docinducedlmprobmulti{\doc'} to variations in the length of $\doc'$.

For a given cluster $\cluster$, 
the
corresponding language model $\clustinducedlmprob{\cdot}$ is induced
by concatenating $\cluster$'s component documents and then applying
the document-LM induction method to the new ``document''.

\paragraph*{Reference comparisons}  While one of our goals is to
demonstrate 
that 
incorporating corpus structure as in our retrieval framework
can provide 
improvements
over the performance of the standard LM algorithm,
we also wish to detemine
whether our algorithms are competitive with 
state-of-the-art language-modeling-based algorithms. One natural
choice for comparison is Lafferty and Zhai's {\em pseudo-feedback
Markov chains} algorithm \cite{Lafferty+Zhai:01a},  which extends the {
expanded query language model} described above by forcing the chains
to pass through top-ranked documents, as determined using the standard
LM approach.  Another obvious candidate is  Lavrenko and Croft's {\em
relevance model} \cite{Lavrenko+Croft:01a} which was the first 
method to explicitly incorporate relevance into the language-modeling
framework, and which demonstrated excellent performance. Note that
both algorithms, in contrast to our framework, depend on
pseudo-feedback mechanisms to cope with the lack of true user feedback.

\paragraph*{Implementation}
We used the Lemur toolkit \cite{Ogilvie+Callan:2001} to run our experiments.  
Our implementations of the 
baseline
used {optimized} smoothing-parameter settings with respect to average
non-interpolated 
precision\footnote{Optimization with respect to recall yielded results
 which were statistically indistinguishable with respect to each 
of 
our performance metrics.}, computed via line search.
For  our novel algorithms, we optimized the cluster-size parameter 
$\sizeclust$ and the \microAlg algorithm's interpolation parameter
 $\lambda$, but
 the other parameters were set to default 
values suggested in the
 previous literature \cite{Zhai+Lafferty:01a}; thus, the baseline 
algorithm was given an
 extra 
advantage. 

Rather than re-implement the pseudo-feedback Markov-chain
and relevance-model
algorithms 
described above, we report 
 results presented in the previous literature \cite{Lafferty+Zhai:01a,Lavrenko+Croft:01a}. We do realize that
 minor differences in performance could stem from specific
 implementation issues, but as stated above, our goal was to test the
 competitiveness of our algorithms' performance with respect to that
 of other prominent algorithms, 
 not 
to prove our algorithms' superiority.

\section{Experimental results}
\label{sec:results}

For our evaluation measures, we used average non-inter\-polated precision, interpolated precision at $0$, and recall,
all for $\numretdocs=1000$ selected documents.
Our main experimental results are given by Tables \ref{tab:AP89},
\ref{tab:AP88+89}, and \ref{tab:LA+FR} and Figure
\ref{fig:interPrec}. 

In the tables, for each evaluation metric, the strongest performance
is boldfaced and all results above the baseline (\baseline) are
italicized. Also, the Wilcoxon two-sided test was employed with
significance threshold $p=0.05$ --- all statistically
significant 
performance improvements and degradations
for our algorithms 
relative to the baseline
 are marked with a star (*).

Clearly, at the indicated settings (given in the captions), 
even at worst our algorithms are always competitive with the 
baseline LM approach, and with occasional exceptions (mostly for
precision at 0) generally do better.  
We also observe that the \mrr and \microAlg algorithms are competitive
with the pseudo-feedback Markov-chains algorithm 
(see Table \ref{tab:AP89}) and the relevance-model algorithm 
(see Table \ref{tab:AP88+89}) with respect to all performance measures.

Figure \ref{fig:interPrec} shows 11-point precision/recall curves for
our algorithms and the baseline.  In all
three corpora, the \microAlg algorithm does best overall.  On
AP88 and AP88+89, 
our cluster-based algorithms on the whole generally perform demonstrably better
than the baseline. In LA+FR, however,  the new
algorithms, with the exception of  the \microAlg algorithm, seem difficult to
distinguish from \baseline, as is borne
out by the 
relative lack of
statistical-significance indications in Table
\ref{tab:LA+FR}.

The fact that the \mrr algorithm was usually superior to
\doubleCountAspect indicates that incorporating within-cluster
structure, as represented by $\clustinducedlmprob{\doc}$, is
important.

Finally, the generally high performance of our \mrr and  \microAlg 
algorithms seems to support our claims as to the importance of using
corpus-structural information in the particular ways we have
suggested: specifically, in these two algorithms, clusters play both a
selection and a
smoothing role, and both document-specific information and intra-cluster
structure are incorporated as well.

\medskip

In what follows, we discuss the results of further experimental
studies.  For space reasons, we present only a subset of the
performance figures for a selection of corpora.

\paragraph*{Parameter selection} 

The cluster-size parameter $\sizeclust$ does have a noticeable impact
on performance. 
A series of 
preliminary experiments (whose results are
omitted 
due to space restrictions)
indicate 
that 
small values of $\sizeclust$ (e.g., 5 or 10) 
yield better results than the baseline LM for all but the
\doubleCountAspect method, 
demonstrating the usefulness of even tiny document clusters. However,
increasing  $\sizeclust$ 
to 40  
resulted in superior performance on the AP89 and AP88+89 datasets,
which 
suggests that the re-rank step of our algorithm
template can compensate to a degree for 
the extra irrelevant documents
that large clusters may bring into consideration.

\begin{figure*}[t]
\begin{tabular}{cc}
{\bf AP89} & {\bf AP88+89} \\
\epsfig{file= 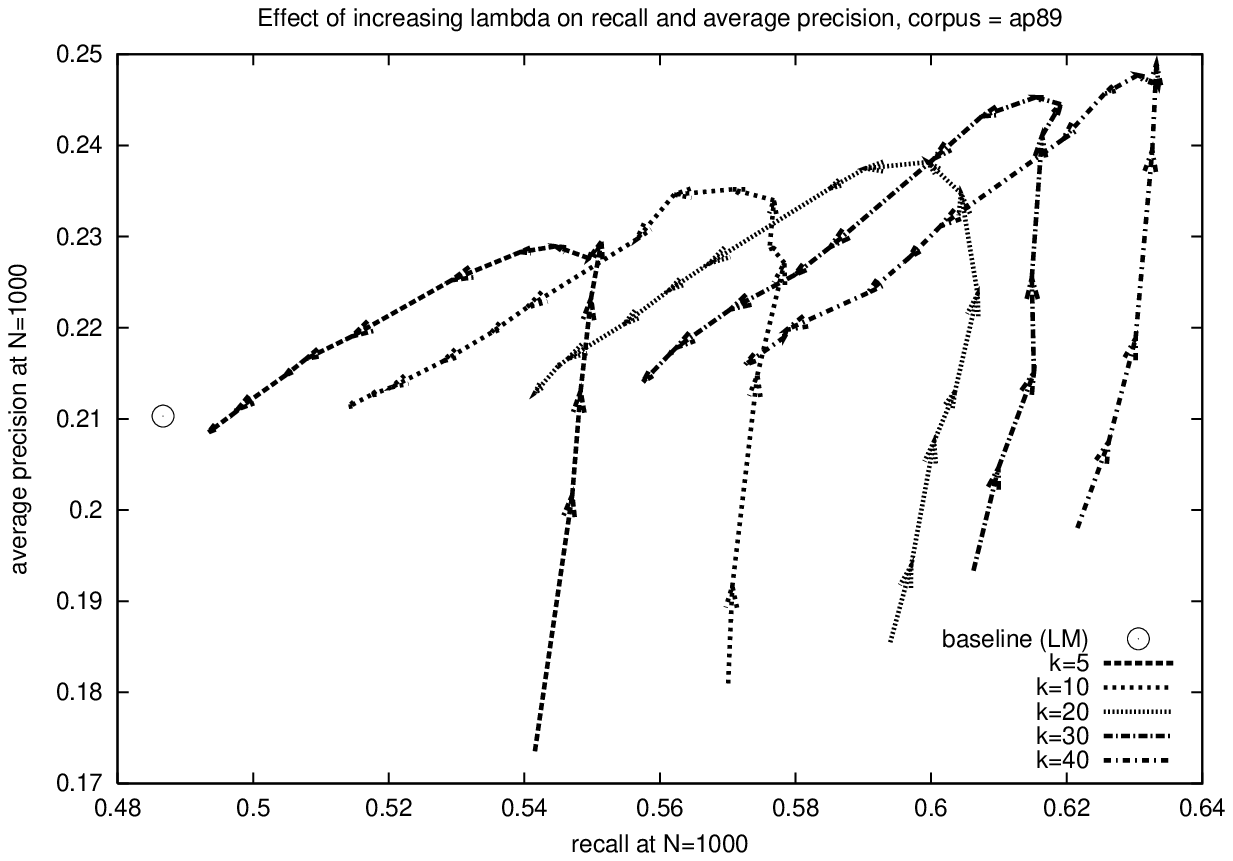,
  width= 3.5in}
&
\epsfig{file=
  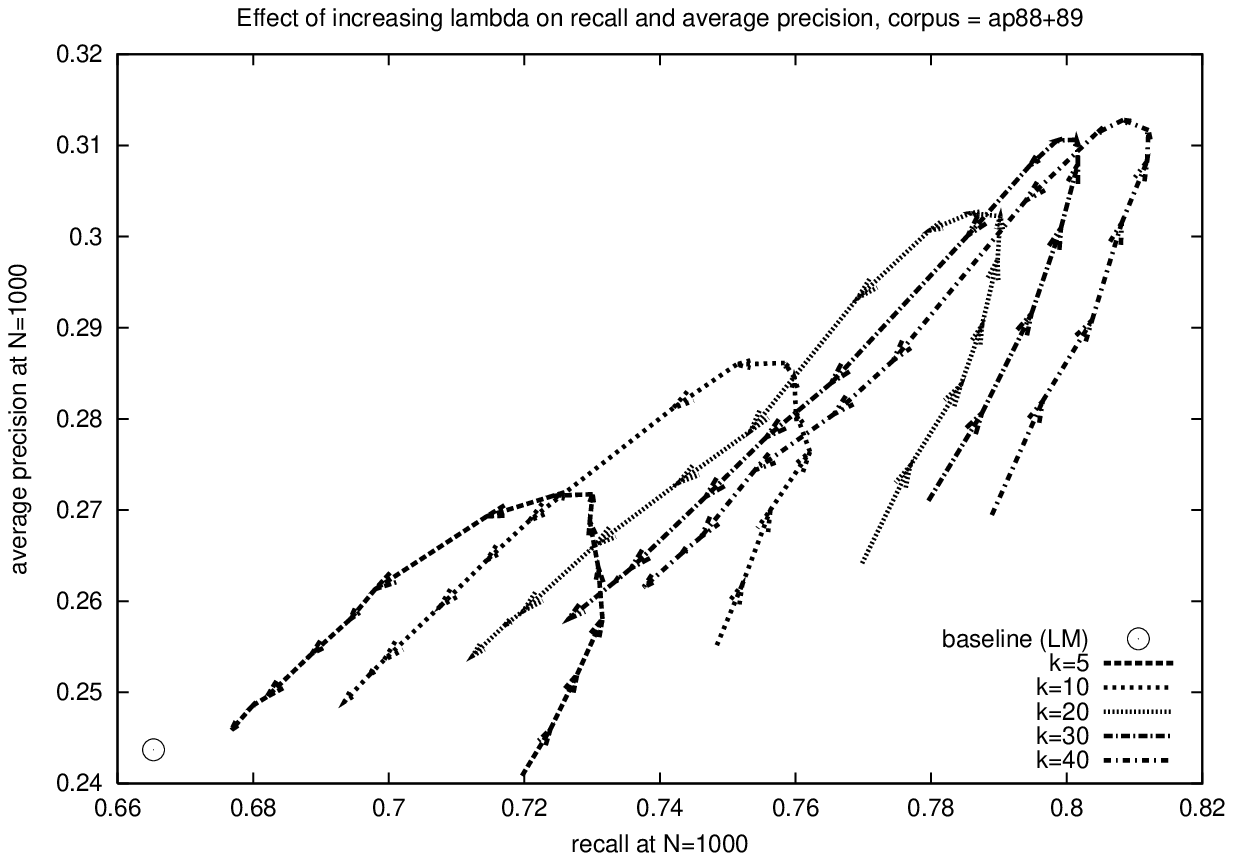, width=
  3.5in }
\end{tabular}
\caption{\label{fig:lambdaVary} \MicroAlg algorithm's recall vs average precision as  $\lambda$ grows
  (increments of .1 until .9, then .925, .95, .975, .98, .99).
  Recall that $\lambda=1$ would yield the baseline language-model scoring
  function.
Similar patterns were observed on the LA+FR corpus; we omit the
  results for clarity.
}

\end{figure*}

We must also choose $\numretclust$, the number of clusters to be
retrieved, recalling that we wish to return a fixed number
$\numretdocs=1000$ of documents.  In the experimental results reported
above, two different schemes were used.  For the algorithms using
clusters solely for selection, we set $\numretclust$ to either 1000 or
the minimum value needed for there to be $1000$ documents receiving a
non-zero score.\footnote{In the \doubleCounted algorithm we chose
$\numretclust=1000$, although lower values would have sufficed.} For
the remaining algorithms (\mrr, \doubleCountAspect, and \microAlg), we
set $\numretclust=10000$.  The former group of algorithms were more
sensitive to choice of $\numretclust$ than the latter, where as long
as $\numretclust$ did not exceed $10000$, 
satisfactory improvements with respect to
the baseline algorithms were observed.
However, drawing upon more clusters than this
--- which in a sense is what the classic aspect model
\cite{Hofmann+Puzicha:98b} does --- was clearly detrimental for some of the data corpora.

An important regard in which the \microAlg algorithm differs from the
other methods we have introduced is 
in its inclusion of an
 additional free parameter
$\lambda$, representing the degree of dependence on
$\docinducedlmprob{\query}$ relative to the \mrr algorithm.  Figure
\ref{fig:lambdaVary} plots the ``trajectories'' of the \microAlg
algorithm through performance space as $\lambda$ is increased. This
figure makes visually clear the interplay between cluster and document
information: small $\lambda$'s (emphasizing clusters) result in better
recall but relatively poor precision; but large $\lambda$'s
(emphasizing documents) 
improve precision at the expense of recall.
The performance of ``average'' values (around .6)
shows that integrating document- and cluster-level information provides
better performance than either can produce alone.

We note that the \mrr algorithm can be viewed as a version of the \microAlg
algorithm in which $\lambda=0$ and re-ranking is added to improve
average precision.  In return for some performance degradation
relative to the \microAlg algorithm, it offers the advantage of having
one fewer parameter to tune, and is fairly robust to 
$\numretclust$'s value as well.

\paragraph*{The re-rank step}
How important is the re-ranking step, in which the top-ranked documents
are re-scored by their document-specific language models, 
to producing good precision?  We ran
several experiments to explore this issue.

First, we observed considerable degradation in average precision if we
removed the $\docinducedlmprob{\query}$ term from the score functions
of the
\centroidAlg and \fifoAlg algorithms, for which re-ranking is redundant.
{Note that this version of the
\centroidAlg algorithm corresponds to applying the basic LM
approach to the ``document'' $\nbhd{\doc}$ rather than $\doc$ itself,
and so can be thought of as a smoothing method wherein the document
language model is created by backing off completely to a cluster language model.}

Next, we examined the role of  the optional re-ranking
step in the algorithms that explicitly incorporate it. 
When the \mrr and
\doubleCountAspect algorithms --- the two cases in which the scoring
function does not incorporate $\docinducedlmprob{\query}$ --- were run
without the optional re-ranking phase,  low average precision and precision at 0
resulted, implying that reliance on $\clustinducedlmprob{\cdot}$ alone
suffers from over-regularization;
the results for the \mrr algorithm are shown in Table \ref{tab:aspectXReRank}.
Furthermore, re-ranking is also required to achieve
reasonable precision for the \doubleCounted algorithm, even though its
scoring function incorporates $\docinducedlmprob{\query}$:  when re-ranking is
not applied,  average precision suffers
  when clusters are small.

\begin{table}
\begin{flushleft}
\scriptsize
\centering
\begin{tabular}{|l|c|c|c|c|c|c|} \hline
& \multicolumn{2}{|c|}  {AP89} & \multicolumn{2}{|c|} {AP88+89} & \multicolumn{2}{|c|} {LA+FR}  \\ \hline

re-rank? & {yes} & {no} & {yes} & {no} & {yes} & {no} \\ \hline
{Avg. Prec.} & $22.6\%$& $19.8\%$ & $27.5\%$& $26.95\%$ & $22.45\%$  & $16.01\%$ \\ \hline
{Prec. at $0$} & $58.2\%$ & $46.98\%$ &  $65.9\%$ & $65.21\%$ &  $58.25\%$ & $46.92\%$ \\ \hline

\end{tabular}
\caption{\label{tab:aspectXReRank} Effect of re-rank step on \mrr precision. For AP89 and AP88+89, k=40; for LA+FR, k=10.} 
\end{flushleft}
\end{table}

\begin{figure}[t]
\newcommand{\mytabwidth}{3.4in}
\begin{centering}
\begin{tabular}{c}
\epsfig{figure= 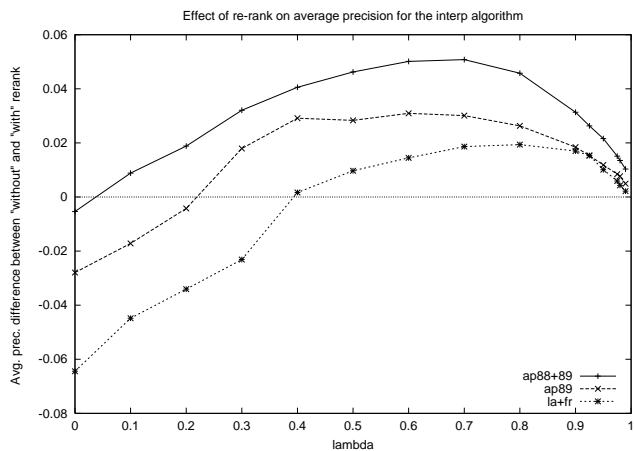 , width=\mytabwidth}
\end{tabular}
\end{centering}
\caption{\label{fig:interpRankeEffect}Effect of re-rank step on
 average precision for the \microAlg algorithm
as $\lambda$ varies. 
For AP89 and AP88+89 k=40; for LA+FR k=10.}
\end{figure}

In the case of the \microAlg algorithm, however, the additional re-rank phase is not
needed as long as the interpolation weight $\lambda$ for the document-based language
model is large enough.
This 
can be seen in Figure \ref{fig:interpRankeEffect}, where the
difference between average precision 
without and with re-rank at different values of $\lambda$ is
 reported --- observe that for $\lambda > 0.4$, re-ranking degrades
 performance.  

These results suggest that (1) for best results, it is important to strike
 the right balance between document-specific and inter-document
 information, and (2) for some algorithms, 
re-ranking creates
 this balance, but in others it can upset it.

\paragraph*{Smoothing}  The sensitivity of the LM approach to choice
of smoothing technique and smoothing parameters has promp\-ted a great
deal of research
\cite{Zhai+Lafferty:01a,Hiemstra:02a,Zaragoza+Hiemstra+Tipping:03a}.
However, we found that for our algorithms, simply setting the
Dirichlet smoothing parameter $\mu$ to a suggested value of 2000
\cite{Zhai+Lafferty:01a} (or, as
it turned out, randomly-chosen values within the neighborhood of 2000)
 outperformed the $\mu$-optimized baseline.
Moreover, experiments with Jelinek-Mercer and 
absolute dis\-counting --- two other well-known 
single-parameter smoothing methods \cite{Zhai+Lafferty:01a} ---
yielded the same outcome
of relative insensitivity to choice of parameter value for the
underlying smoothing method employed.

\paragraph*{Feature selection}
Another interesting observation is that effective incorporation of
cluster information somewhat obviates the need for feature selection.
In particular, Table \ref{tab:AP89UnStemmed} shows one case where
using the \mrr and \microAlg algorithms {\em without} a stemmer or
stop-word list outperforms the baseline {\em with} access to the Porter
stemmer with respect to average
precision and recall.
On the other hand, stemming led to degradation of precision at zero.
Results for cluster sizes other than 40 and different corpora
were consistent with these findings.

\begin{table*}[t]
\begin{flushleft}
\small
\centering
\begin{tabular}{|l|c|c||c|c|c|c|c|}
\hline
 &S-Baseline& U-Baseline& S-\mrr& U-\mrr& S-\microAlg ($\lambda=0.4$)&
 U-\microAlg ($\lambda=0.6$) \\ \hline \hline
Avg. Prec.& $21.03\%$ & $19.56\%$ & $\better{22.6\%^*}$ & $\better{21.51\%^*}$ & \mbox{\boldmath$24.9\%^*$} & $\better{24.08\%^*}$ \\ \hline
Prec. at 0 & $57.4\%$ & $60.1\%$ & $\better{58.2\%}$ &  \mbox{\boldmath$60.9\%$} & $55.8\%$ & $58.9\%$\\ \hline
\relret{RelRet & $1587/3261$ & $1467/3261$ & $\better{2027/3261}$ &  $\better{1972/3261}$ & \mbox{\boldmath$2065/3261$} & $\better{1978/3261}$ \\ \hline}
Recall & $48.67\%$ & $44.99\%$ & $\better{62.16\%^*}$  & $\better{60.47\%^*}$ & \mbox{\boldmath$63.62\%^*$} & $\better{60.66\%^*}$ \\ \hline
\end{tabular}
\caption{\label{tab:AP89UnStemmed} Stemming comparison on AP89. 
S-: stemmed version; U-: un-stemmed version. 
Cluster size $\sizeclust=
40$.  Significant differences are reported with respect to the corresponding baseline.
}
\end{flushleft}
\end{table*}

\smallskip
\noindent{\subsecfnt{Is it all due to language modeling?}}
Throughout this paper, we have 
used language models as our information representation.  
An interesting question is whether it is the representation (e.g., $\clustinducedlmprob{\cdot}$), or
the {\em source} of this representation (e.g. $c$ itself) that matters most.
We therefore explored the effect of using an alternative
representation.  Specifically, both the queries and the documents were
represented using log-based tf.idf, with the inner product as distance
measure.  As before, clusters were treated as large documents formed
by concatenating their contents.
Altering our 
selection 
algorithms (\centroidAlg,
\fifoAlg, and \doubleCounted )
in this way led to improved performance with respect to the  basic
tf.idf retrieval algorithm, as shown in Table \ref{tab:LA+FRtfidf}. 
On the other hand, these algorithms did not do as well as their
original,
LM-based counterparts.
We thus see that our algorithmic framework can boost performance for
other information representations over the structure-blind
alternative, but language models do seem to have advantages, at least
in comparison to tf.idf.

\begin{table*}[t]
\begin{flushleft}
\small
\newcommand{\tfpref}{}
\newcommand{\lmpref}{}
\centering
\begin{tabular}{|l|c|c|c|c||c|c|c|c|}
\hline

 & \multicolumn{4}{|c||}{tf.idf version} & \multicolumn{4}{c|}{LM
 version} \\\cline{2-9}
 & {\tfpref}Baseline & {\tfpref}\centroidAlg  & {\tfpref}\fifoAlg & {\tfpref}\doubleCounted & {\lmpref}Baseline & {\lmpref}\centroidAlg & {\lmpref}\fifoAlg & {\lmpref}\doubleCounted \\  \hline \hline
Avg. Prec.& $16.43\%$ &  $\better{16.67\%}$ & \mbox{\boldmath$17.18\%^*$} & $\better{16.68\%^*}$& ${22.16\%}$ & ${21.92\%}$ & \mbox{\boldmath$22.52\%$} &  ${22\%}$ \\ \hline
Prec. at 0 & $46.66\%$ &$\better{46.94\%}$ & \mbox{\boldmath$47.29\%$} & $\better{46.92\%}$& ${57.37\%}$ & $\better{57.91\%}$ & $\better{57.89\%}$ & \mbox{\boldmath$58.16\%$}  \\ \hline
\relret{RelRet & $660\relthree$ & $\better{766\relthree}$ & \mbox{\boldmath$801\relthree$} & $\better{723\relthree}$ & ${674\relthree}$ & $\better{774\relthree}$ & \mbox{\boldmath$808\relthree$}& $\better{742\relthree}$   \\ \hline}
Recall & $47.45\%$ & $\better{55.07\%}$ & \mbox{\boldmath$57.58\%$} & $\better{51.98\%}$ & ${48.31\%}$ & $\better{55.64\%}$ & \mbox{\boldmath$58.09\%$} &$\better{53.34\%}$  \\ \hline
\end{tabular}
\caption{\label{tab:LA+FRtfidf} 
Simple similarity metric based on tf.idf vs. LM-based
similarity on  LA+FR.  Cluster size $\sizeclust=10$.
}
\end{flushleft}
\end{table*}

\section{Conclusions}

In summary, we have proposed a general framework that enables the
development of a variety of algorithms 
for integrating corpus similarity structure, modeled via clusters,
and document-specific information. 
Although our proposal is motivated by the recent 
language-modeling approach to information retrieval, and the specific
 algorithms presented here do use language models for representation
purposes to good effect, 
we observed that 
the
framework  also can be used 
with basic 
classic IR techniques such as tf.idf.

An interesting direction for future work is to explore
the effect of using alternative clustering algorithms.  We would also
like to study 
the role that overlapping plays in our framework: is most of the
performance gain due to the (high) degree of overlap in our clusters
or to the way structure and 
individual-document information are
integrated?
Another interesting direction is to 
examine whether other
algorithms, such as the  LM-based pseudo-feedback  methods
we used for reference comparisons
\cite{Lafferty+Zhai:01a,Lavrenko+Croft:01a},  can benefit if we
replace the basic LM retrieval algorithm 
they employ
with 
one of ours.

Most importantly, we would like to develop a principled probabilistic
interpretation of the framework we have proposed.  We have done some
preliminary work based on considering the factorization
$\condP{\query}{\doc} = \sum_\cluster
\condP{\query}{\doc,\cluster}\condP{\cluster}{\doc}$; some of the
components of our scoring functions can be considered to be (very
rough) approximations of the terms in this factorization. Creating a
rigorous probabilistic foundation 
for the work described here is one of
our main future goals.

\medskip

\noindent {\bf Acknowledgments} ~ We thank Eric Breck, Claire Cardie,  Shimon Edelman,
Thor\-sten Jo\-ach\-ims, Art Munson, Bo Pang, Ves Stoyanov, and the
anonymous reviewers for valuable comments.  Thanks to 
ChengXiang Zhai
and Victor Lavrenko for answering questions about their work,
and Andr\'es Cor\-ra\-da-Emmanuel for 
responding to queries about Lemur.
This paper is based upon work supported in part by the National
Science Foundation under grants ITR/IM IIS-0081334 and IIS-0329064 and
by an Alfred P. Sloan Research Fellowship. Any opinions, findings, and
conclusions or recommendations expressed above are those of the
authors and do not necessarily reflect the views of the National
Science Foundation or Sloan Foundation.

\end{document}